\documentclass[preprint]{emulateapj}
\usepackage{graphicx}
\usepackage{amsmath}
\usepackage{epstopdf}
\usepackage{amssymb}
\usepackage{extarrows}
\usepackage{color}
\usepackage{bm}
\usepackage{natbib}

\newcommand{\be}{\begin{equation}}
\newcommand{\ee}{\end{equation}}
\newcommand{\ba}{\begin{eqnarray}}
\newcommand{\ea}{\end{eqnarray}}

\newcommand{\au}{\mathrm{AU}}
\newcommand{\IN}{\mathrm{in}}
\newcommand{\OUT}{\mathrm{out}}

\newcommand{\lk}{\mathrm{LK}}

\newcommand{\eff}{\mathrm{eff}}

\newcommand{\m}{\mathrm{max}}
\newcommand{\mi}{\mathrm{min}}
\newcommand{\li}{\mathrm{lim}}

\newcommand{\gr}{\mathrm{GR}}
\newcommand{\f}{\mathrm{f}}
\newcommand{\SL}{\mathrm{SL}}

\begin{document}
\title{Spin-Orbit Misalignment of Merging Black-Hole Binaries
       with Tertiary Companions}
\author{Bin Liu$^{1,2}$ and Dong Lai$^{2}$}
\affiliation{$^{1}$ Shanghai Astronomical Observatory, Chinese Academy of Sciences, 80 Nandan Road, Shanghai 200030, China}
\affiliation{$^{2}$ Cornell Center for Astrophysics and Planetary Science, Cornell University, Ithaca, NY 14853, USA}

\begin{abstract}
We study the effect of external companion on the orbital and spin
evolution of merging black-hole (BH) binaries. An sufficiently close
by and inclined companion can excite Lidov-Kozai (LK) eccentricity
oscillations in the binary, thereby shortening its merger time. During
such LK-enhanced orbital decay, the spin axis of the BH generally
exhibits chaotic evolution, leading to a wide range ($0^\circ$-$180^\circ$) of
final spin-orbit misalignment angle from an initially aligned
configuration. For systems that do not experience eccentricity
excitation, only modest ($\lesssim 20^\circ$) spin-orbit misalignment
can be produced, and we derive an analytic expression for the final
misalignment using the principle of adiabatic invariance.  The
spin-orbit misalignment directly impacts the gravitational waveform, and
can be used to constrain the formation scenarios of BH binaries and
dynamical influences of external companions.
\end{abstract}

\maketitle

\section{Introduction}

The recent breakthrough in the detection of
gravitational waves (GWs) from merging black hole (BH) binaries by advanced LIGO
\citep{Abbott 2016a,Abbott 2016b,Abbott 2017} has
generated renewed interest in understanding the formation mechanisms of compact
BH binaries,
from the evolution of massive stellar binaries
\citep{Lipunov 1997,Belczynski 2010,Belczynski 2016,Mandel and de Mink
  2016,Lipunov 2017} and triples \citep{Silsbee and Tremaine
  2017,Antonini 2017} in the galactic fields, to dynamical
interactions in galactic nuclei \citep{Antonini 2012,Petrovich 2017}
and in the dense core of globular clusters \citep{Miller 2002,Rodriguez
  2015,Chatterjee 2017}.

Because of the uncertainties associated with various formation channels
(e.g., common envelope evolution in the standard binary channel),
it is difficult to distinguish the different formation
mechanisms based on BH mass measurement alone. The detection of
eccentric systems would obviously indicates some dynamical processes at
work (e.g., \citet{Antonini 2012,Silsbee and Tremaine 2017}).  However,
because of the efficient eccentricity damping by gravitational
radiation, the vast majority of compact binaries will likely be
circular when entering the LIGO sensitivity band regardless of the
formation channels.  It has been suggested that the BH spin and the
spin-orbit misalignment angle may be an important discriminant. In
particular, the spin-orbit misalignment directly impacts the projected
spin parameter of the merging binaries,
\be \chi_{\rm eff}={m_1 {\bf a}_1 + m_2{\bf a}_2\over m_1+m_2}\cdot \hat{\textbf{L}},
\ee
(where $m_{1,2}$ are the BH masses, ${\bf a}_{1,2}=c{\bf S}_{1,2}/(Gm_{1,2}^2)$
are the dimensionless BH spins, and $\hat{\textbf{L}}$ is the unit orbital angular momentum vector),
which can be measured from the phase evolution of GWs \citep{Abbott 2016b,Abbott 2017}.

In this paper, we study the merger and spin-orbit misalignment of BH binaries
in the presence of tertiary companion. Such triple BH systems could be
a direct product of massive triple stars in the galactic field \citep{Silsbee and Tremaine 2017,Antonini 2017},
or could be produced dynamically in a dense cluster \citep{Miller 2002,Rodriguez 2015,Antonini and Rasio 2016}.
For binaries formed near the center of a galaxy, the third body could be a supermassive BH
\citep{Antonini 2012,Petrovich 2017}.

It is well known that a tertiary body on an inclined orbit can
accelerate the decay of an inner binary by inducing Lidov-Kozai (LK)
eccentrcity/inclination oscillations \citep{Lidov,Kozai}. This has been
studied before in the contexts of supermassive BH binary merger
\citep{Blaes 2002} and stellar mass BH binaries
(e.g., \citet{Miller 2002,Thompson 2011,Antonini 2014,Silsbee and Tremaine 2017}).
We focus on the latter in this paper.
We show that as the BH binary undergoes LK-enhanced decay from a wide orbit
and eventually enters the LIGO band, the spin axis of individual BHs
can experience chaotic evolution, so that a significant spin-orbit misalignment
can be produced prior to merger even for binaries formed with zero initial misalignment.
We derive relevant analytic relations and
quantify how the final spin-orbit misalignment angle
depends on various parameters of the system (binary and external companion).

Note that in this paper we focus on triple systems with relatively
small binary separations ($\lesssim 0.2$~AU for the inner binaries),
so that the inner binary can merger within $10^{10}$~years either by
itself or through modest ($e\lesssim 0.99$) LK eccentricity excitation.
Such compact triple systems likely have gone through a complex (and
highly uncertain) sequence of common envelop or mass transfer
evolution -- we do not study such evolution in this paper and therefore
do not address issues related to the occurrence rate of compact triples.
Our goal is to use such triple systems to illustrate the complex spin dynamics
of the individual BHs. We expect that similar spin dynamics may take place
in other types of triple systems, e.g., those with much	larger initial separations,
but experience extreme eccentricity excitation due	to non-secular forcing	
from tertiary companions \citep{Silsbee and Tremaine 2017,Antonini 2017}.

\section{Lidov-Kozai cycles in BH triples with gravitational radiation}
\subsection{Setup and Orbital Evolution}

Consider a hierarchical triple system, consisting of
an inner BH binary with masses $m_1$, $m_2$
and a relatively distant companion of mass $m_3$.
The reduced mass for the inner binary is $\mu_\IN\equiv m_1m_2/m_{12}$, with $m_{12}\equiv m_1+m_2$.
Similarly, the outer binary has $\mu_\OUT\equiv(m_{12}m_3)/m_{123}$ with $m_{123}\equiv m_{12}+m_3$.
The orbital semimajor axes and eccentricities are denoted by $a_{\IN,\OUT}$ and $e_{\IN,\OUT}$, respectively.
The orbital angular momenta of the inner and outer binaries are
\begin{eqnarray}
&&\textbf{L}_\IN=\mathrm{L}_\IN\hat{\textbf{L}}_\IN=\mu_\IN\sqrt{G m_{12}a_\IN(1-e_\IN^2)}\,\hat{\textbf{L}}_\IN,\\
&&\textbf{L}_\OUT=\mathrm{L}_\OUT\hat{\textbf{L}}_\OUT=\mu_\OUT\sqrt{G m_{123}a_\OUT(1-e_\OUT^2)}\,\hat{\textbf{L}}_\OUT,
\end{eqnarray}
where $\hat{\textbf{L}}_{\IN,\OUT}$ are unit vectors. The
relative inclination between $\hat{\bf L}_\IN$ and $\hat{\bf L}_\OUT$ is denoted by $I$.
For convenience, we will frequently omit the subscript ``$\IN$".

The merger time due to GW radiation of an isolated binary with initial $a_0$ and $e_0=0$
is given by
\be
\begin{split}
T_\mathrm{m,0}&=\frac{5c^5 a_0^4}{256 G^3 m_{12}^2 \mu}\\
&\simeq 10^{10}\bigg(\frac{60M_\odot}{m_{12}}\bigg)^2\bigg(\frac{15M_\odot}{\mu}\bigg)\bigg(\frac{a_0}{0.202\au}
\bigg)^4{\rm yrs}.
\end{split}
\label{eq:tm0}\ee
A sufficiently inclined external companion can raise the binary eccentricity through Lidov-Kozai
oscillations, thereby reducing the merger time or making
an otherwise non-merging binary merge within $10^{10}$~years.
To study the evolution of merging BH binary under the influence of a companion,
we use the secular equations to the octupole level in terms of the angular momentum
$\textbf{L}$ and eccentricity $\textbf{e}$ vectors:
\begin{eqnarray}
&&\frac{d \textbf{L}}{dt}=\frac{d \textbf{L}}{dt}\bigg|_{\mathrm{LK}}+\frac{d \textbf{L}}{dt}\bigg|_{\mathrm{GW}}~,\label{eq:Full Kozai 1}\\
&&\frac{d \textbf{e}}{dt}=\frac{d \textbf{e}}{dt}\bigg|_{\mathrm{LK}}+\frac{d \textbf{e}}{dt}\bigg|_{\mathrm{GR}}+\frac{d \textbf{e}}{dt}\bigg|_{\mathrm{GW}}~.\label{eq:Full Kozai 2}
\end{eqnarray}
Here the ``Lidov-Kozai'' (LK) terms are given explicitly in \citep{Liu et al 2015}
(we also evolve $\textbf{L}_\OUT$ and $\textbf{e}_\OUT$), and the associated timescale of LK
oscillation is
\be
t_\lk=\frac{1}{n}\frac{m_{12}}{m_3}\bigg(\frac{a_{\OUT,\eff}}{a}\bigg)^3,
\ee
where $n=\sqrt{G m_{12}/a^3}$ is the mean motion of the inner binary
and $a_{\OUT,\eff}\equiv a_\OUT\sqrt{1-e^2_\OUT}$.
General Relativity (1-PN correction) induces pericenter precession
\be
\frac{d \textbf{e}}{dt}\bigg|_{\mathrm{GR}}=\Omega_\mathrm{GR}\hat{\textbf{L}}\times\textbf{e},
~~~\Omega_\mathrm{GR}=\frac{3Gn m_{12}}{c^2a(1-e^2)}.
\ee
We include GW emission (2.5-PN effect) that causes orbital decay and circularization \citep{Shapiro 1983},
but not the extreme eccentricity excitation due to non-secular effects \citep{Antonini 2014,Silsbee and Tremaine 2017}.

\begin{figure}
\begin{centering}
\begin{tabular}{cc}
\includegraphics[width=8cm]{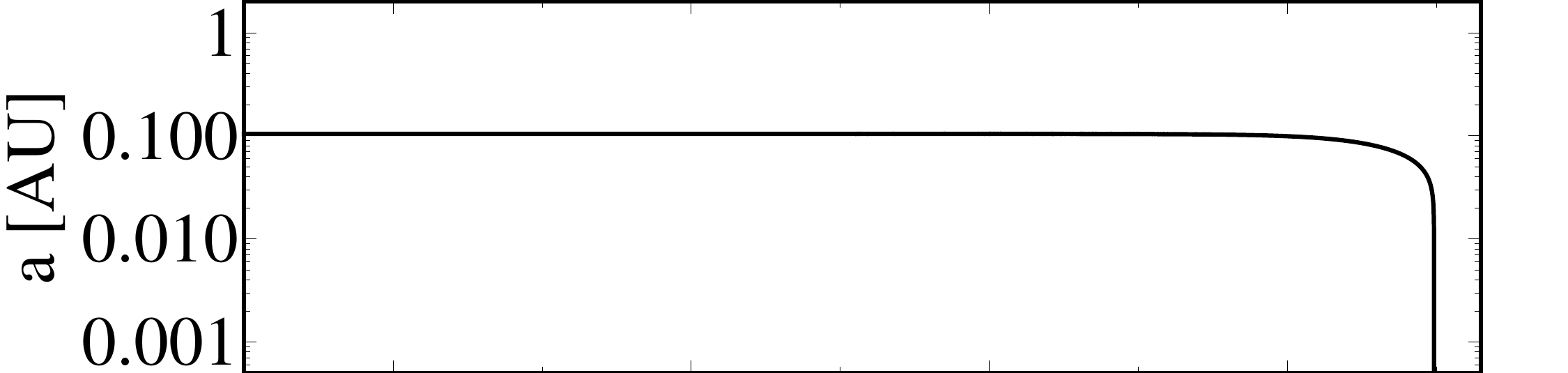}\\
\includegraphics[width=8cm]{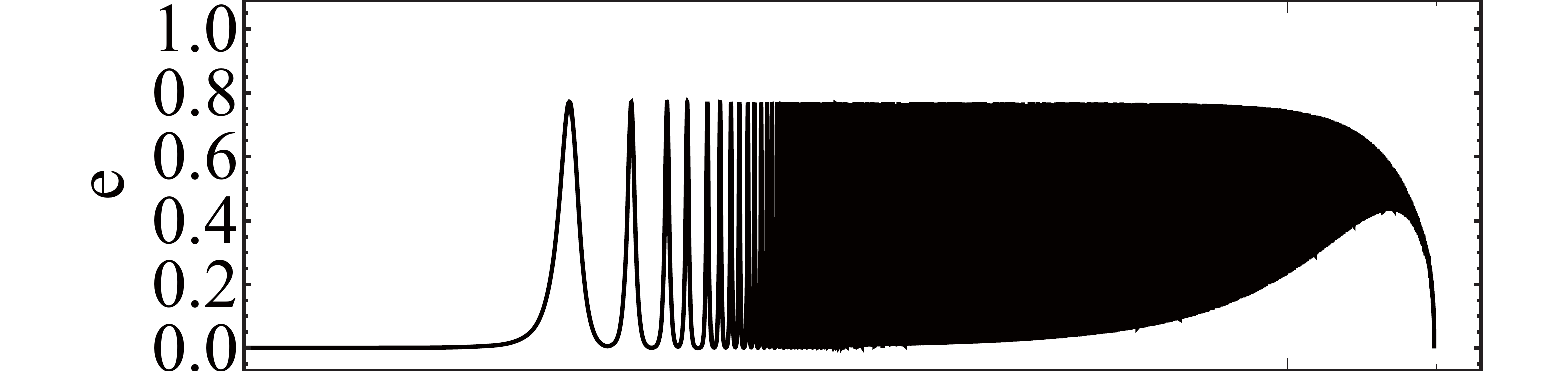}\\
\includegraphics[width=8cm]{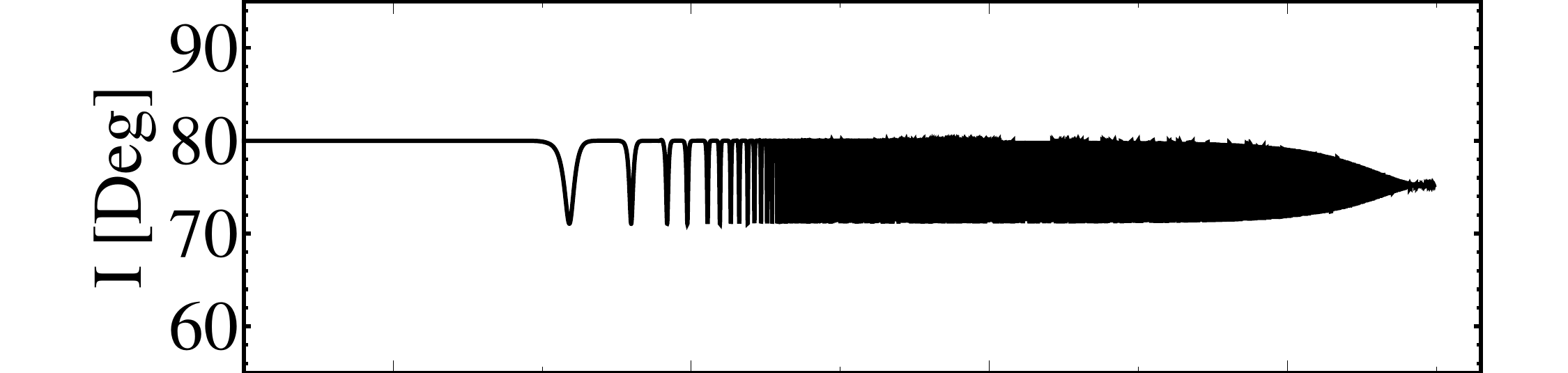}\\
\includegraphics[width=8cm]{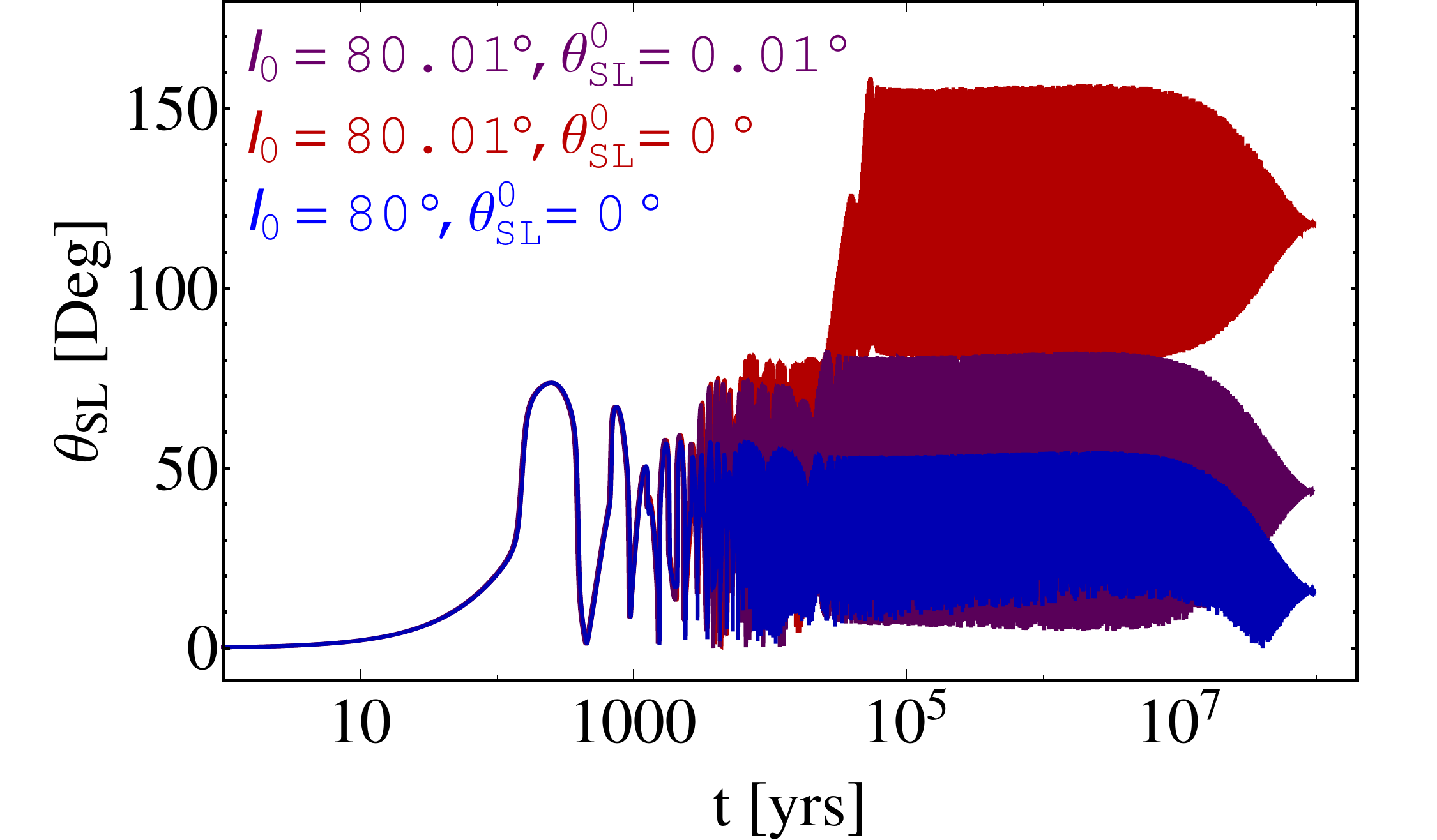}
\end{tabular}
\caption{Sample orbital and spin evolution of a BH binary system with a tertiary companion.
The top three panels show the semimajor axis, eccentricity
and inclination (relative to ${\hat{\bf L}}_\OUT$) of the inner BH binary, and the bottom panel shows
the spin-orbit misalignment (the angle between ${\bf S}_1$ and ${\bf L}$).
The parameters are $m_1=m_2=m_3=30M_\odot$, $a_\OUT=3\au$, $e_\OUT=0$, and the initial
$a_0=0.1\au$, $e_0=0.001$, $I_0=80^\circ$ and $\theta_\SL^0=0^\circ$.
For this example, the octupole effect is absent.
In the bottom panel, the results for slightly different values of $I_0$ and $\theta_\SL^0$ (as indicated)
are plotted, showing a strong dependence of the final $\theta_{\rm SL}$ on the initial conditions.
}
\label{fig:1}
\end{centering}
\end{figure}

The top three panels of Figure~\ref{fig:1} show an example of the orbital evolution of a BH binary
with an inclined companion (initial $I_0=80^\circ$).
We see that the inner binary undergoes cyclic excursions to maximum eccentricity $e_\m$, with
accompanying oscillations in the inclination $I$.
As the binary decays, the range of eccentricity oscillations shrinks.
Eventually the oscillations freeze and the binary experiences ``pure"
orbital decay/circularization governed by GW dissipation.

\subsection{Eccentricity Excitation and Merger Time}

In the quadrupole approximation, the maximum eccentricity
$e_\m$ attained in the LK oscillations (starting from an initial $I_0$ and $e_0\simeq 0$)
can be calculated analytically using energy and angular momentum conservation, according to the equation
\citep{Anderson et al 2017}
\ba
&&\!\!\!\frac{3}{8}\frac{j^2_\mi-1}{j^2_\mi}\bigg[5\left(\cos I_0+\frac{\eta}{2}\right)^2-
\Bigl(3+4\eta\cos I_0+\frac{9}{4}\eta^2\Bigr)j^2_\mi \nonumber\\
&&\quad +\eta^2j^4_\mi\bigg]+\varepsilon_\gr \left(1-j_\mi^{-1}\right)=0,
\label{eq:EMAX}
\ea
where $j_\mi\equiv\sqrt{1-e_\m^2}$, $\eta\equiv(\mathrm{L}/\mathrm{L}_\OUT)_{e=0}$
and $\varepsilon_\gr$ is given by
\ba\label{eq:epsilonGR}
&&\varepsilon_\gr= t_{\rm LK}\Omega_{\rm GR}\Bigr |_{e=0}= \frac{3Gm_{12}^2a_{\OUT,\eff}^3}{c^2a^4m_3}\nonumber\\
&&\simeq 0.96\bigg(\!\frac{m_{12}}{60M_\odot}\!\bigg)^{\!\!2}\!\bigg(\!\frac{m_3}{30M_\odot}\!\bigg)^{\!\!-1}
\!\bigg(\!\frac{a_{\OUT,\eff}}{3\au}\!\bigg)^{\!\!3}\!\bigg(\!\frac{a}{0.1\au}\!\bigg)^{\!\!-4}.
\ea
Note that in the limit of $\eta\rightarrow0$ and $\varepsilon_\gr\rightarrow 0$,
Equation~(\ref{eq:EMAX}) yields the well-known relation $e_\m=\sqrt{1-(5/3)\cos^2 I_0}$.
The maximum possible $e_\m$ for all values of $I_0$, called
$e_\li$, is given by
\be\label{eq:ELIM}
\frac{3}{8}(j_\li^2-1)\left[-3+\frac{\eta^2}{4}\left(\frac{4}{5}j_\li^2-1\right)\right]+
\varepsilon_\gr \left(1-j_\li^{-1}\right)=0,
\ee
and is reached at $\cos I_0=(\eta/10)(4j_\li^2-5)$.
Eccentricity excitation ($e_\m\ge 0$) occurs within a window of inclinations
$(\cos I_0)_-\leqslant\cos I_0\leqslant(\cos I_0)_+$, where \citep{Anderson et al 2017}
\be\label{eq:Kozai window I}
(\cos I_0)_\pm=\frac{1}{10}\Big(-\eta\pm\sqrt{\eta^2+60-\frac{80}{3}\varepsilon_\gr}\Big).
\ee
This window vanishes when
\be
\varepsilon_\gr\ge \frac{9}{4}+\frac{3}{80}\eta^2\qquad ({\rm no~eccentricity~excitation}).
\label{eq:grlim}\ee

\begin{figure}
\begin{centering}
\begin{tabular}{cc}
\includegraphics[width=7.5cm]{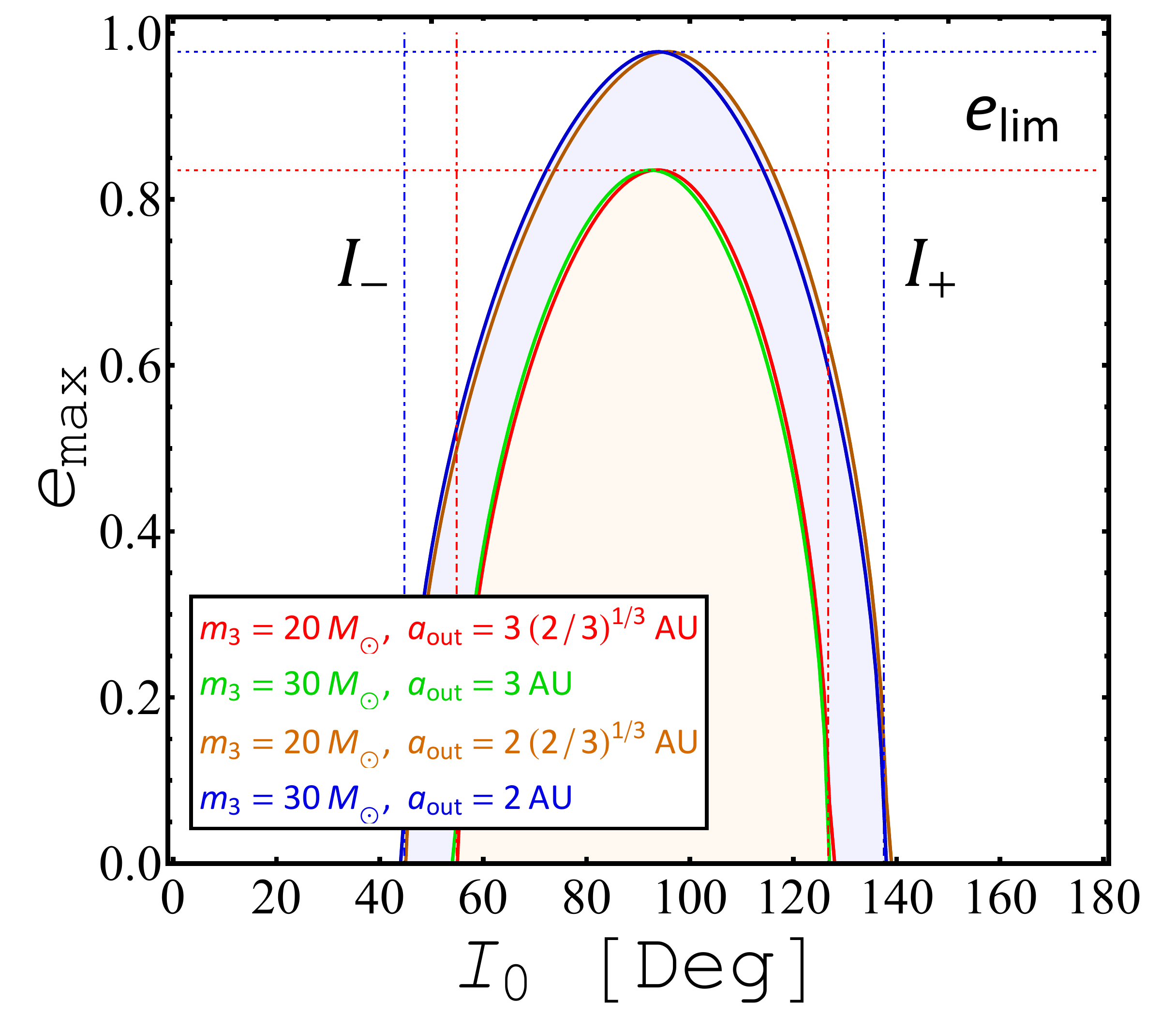}
\end{tabular}
\caption{The maximum eccentricity of the inner BH binary versus the initial inclination $I_0$ of the
tertiary companion, calculated using Equation~(\ref{eq:EMAX}).
The inner binary has $m_1=m_2=30M_\odot$, $a=0.1\au$, and initial $e_0\simeq 0$.
The companion has a circular orbit and its mass and semimajor axis are as labeled.
The $e_{\rm max}(I_0)$ curve depends mainly on $m_3/a_\OUT^3$.
The horizontal ($e_\li$) and vertical ($I_\pm$) lines are given by Equations~(\ref{eq:ELIM})
and (\ref{eq:Kozai window I}), respectively.
}
\label{fig:2}
\end{centering}
\end{figure}

Figure \ref{fig:2} shows some examples of the $e_\m(I_0)$ curves. For $\eta\lesssim 1$,
these curves depend mainly on $m_3/a_{\OUT,\eff}^3$ (for given inner binary parameters).
As $\varepsilon_\gr$ increases (with decreasing $m_3/a_{\OUT,\eff}^3$),
the LK window shrinks and $e_\li$ decreases. Eccentricity excitation is suppressed (for all
$I_0$'s) when Equation~(\ref{eq:grlim}) is satisfied.

For systems with $m_1\neq m_2$ and $e_\OUT\neq 0$, so that
\be
\varepsilon_{\rm oct}\equiv {m_1-m_2\over m_{12}}\left({a\over a_\OUT}\right)
{e_\OUT\over 1-e_\OUT^2}
\ee
is non-negligible, the octupole effect may become important (e.g. \citet{Ford,Naoz 2016}).
This tends to widen the inclination window for large eccentricity excitation.
However, the analytic expression for $e_\li$ given by
Equation~(\ref{eq:ELIM}) remains valid even for $\varepsilon_{\rm oct}\neq 0$ \citep{Liu et al 2015,
Anderson 2017b}. Therefore Equation~(\ref{eq:grlim}) still provides a good criterion
for ``negligible eccentricity excitation''
(note that when $\varepsilon_{\rm oct}\neq 0$, the inner binary cannot be exactly circular;
e.g., \citet{Anderson 2017b,Liu-PRD}).

\begin{figure}
\begin{centering}
\begin{tabular}{cc}
\includegraphics[width=7.5cm]{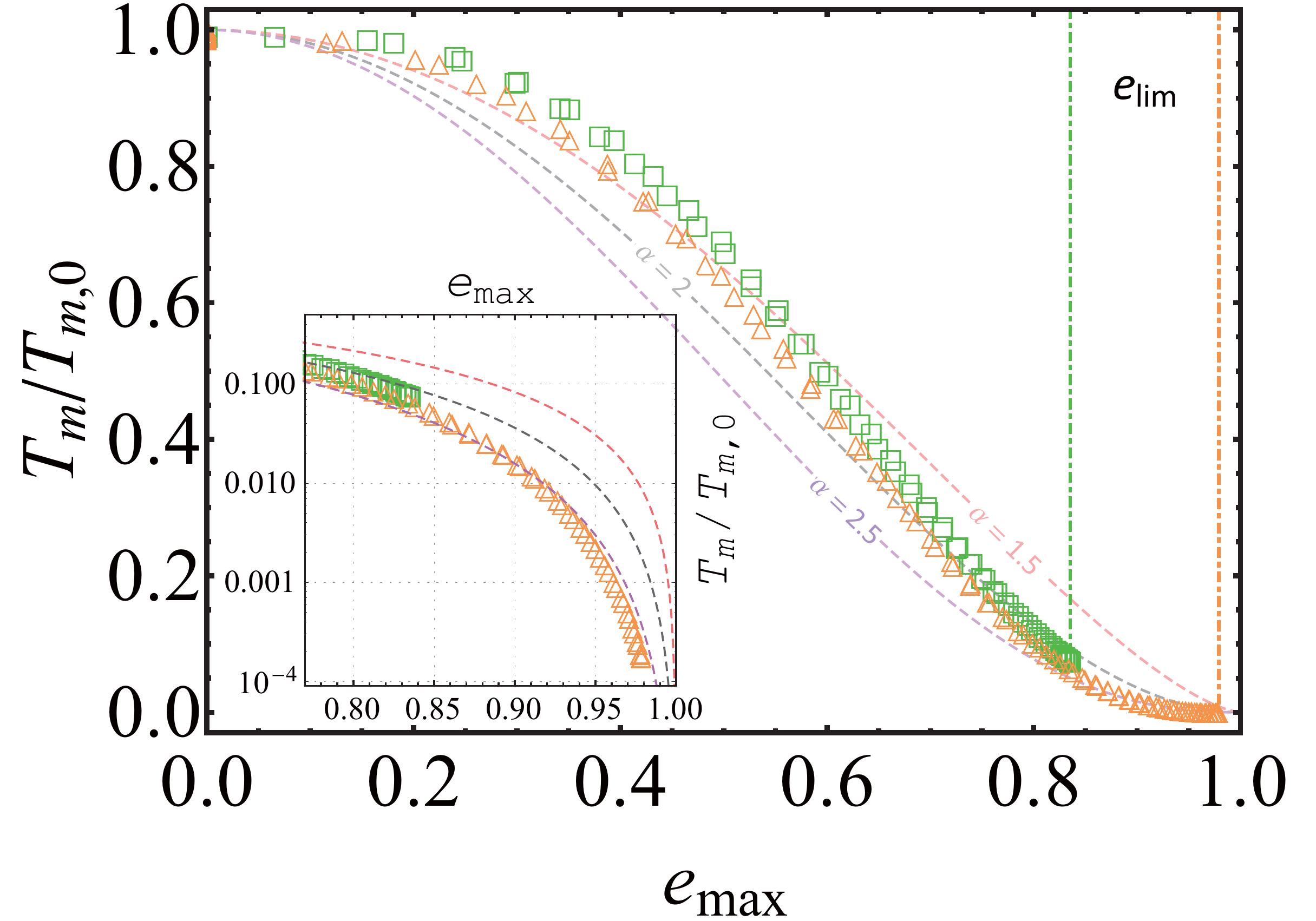}
\end{tabular}
\caption{The binary merger time $T_\mathrm{m}$ (in units of $T_{\mathrm {m},0}$;
Equation~\ref{eq:tm0}) as a function of the maximum eccentricity induced by
a tertiary companion. The BH masses are
$m_1=m_2=m_3=30M_\odot$, and the inner and outer orbits are initially circular.
We consider two sets of initial semimajor axes:
$a_0=0.1\au$, $a_\OUT=3\au$ (orange) and $a_0=0.2\au$, $a_\OUT=5\au$
(green). Here $T_\mathrm {m}$ is computed numerically by
integrating the secular evolution equations with different $I_0$'s,
and $e_\m$ is calculated using Equation~(\ref{eq:EMAX}).  The decreasing
trend of $T_\mathrm{m}/T_\mathrm{m,0}$ as a function of $e_\m$ can
be approximated by $(1-e_\m^2)^\alpha$ (dashed lines).
The inset shows a zoomed-in portion of large $e_\m$'s.}
\label{fig:3}
\end{centering}
\end{figure}

Eccentricity excitation leads to a shorter binary merger time $T_\mathrm{m}$
compared to the circular merger time $T_{\mathrm{m},0}$
(Figure~\ref{fig:3}).  We compute $T_\mathrm{m}$ by integrating
the secular evolution equations of the BH triples with a range of
$I_0$. Each $I_0$ run has a corresponding $e_\m$, which can be
calculated using Equation~(\ref{eq:EMAX}). We see that in general
$T_\mathrm{m}/T_{\mathrm{m},0}$ can be approximated by $(1-e_\m^2)^\alpha$, with
$\alpha\simeq 1.5,~2$ and 2.5 for $e_\m=(0,0.6),~(0.6,0.8)$ and
$(0.8,0.95)$, respectively.

\section{Spin-Orbit dynamics in merging BH binary with an external companion}
\subsection{Spin-Orbit Coupling}

We now study how the BH spin evolves
during the binary merger, considering only
$\textbf{S}_1=\mathrm{S_1} \hat{\textbf{S}}_1$ (where
$\hat{\textbf{S}}_1$ is the unit vector).
The de Sitter precession of $\hat{\textbf{S}}_1$ around $\hat{\textbf{L}}$ is \citep{Barker}
\be\label{eq:spin}
\frac{d \hat{\textbf{S}}_1}{dt}=\Omega_\mathrm{dS}\hat{\textbf{L}} \times \hat{\textbf{S}}_1,
~~~\Omega_\mathrm{dS}=\frac{3 G n (m_2+\mu/3)}{2 c^2 a (1-e^2)}.
\ee
Note that there is a back-reaction torque from ${\bf S}_1$ on ${\bf L}$;
this can be safely neglected since $\mathrm{L}\gg \mathrm{S}_1$.

Before presenting our numerical results, it is useful to note
the different regimes for the evolution of the spin-orbit misalignment angle
$\theta_\SL$ (the angle between ${\bf S}_1$ and ${\bf L}$).
In general, the inner binary axis $\hat{\bf L}$ precesses (and nutates when $e\ne 0$)
around the total angular momentum ${\bf J}={\bf L}+{\bf L}_\OUT$ (recall that $\mathrm{S}_1, \mathrm{S}_2\ll \mathrm{L}$, and
${\bf J}$ is constant in the absence of GW dissipation). The related precession rate
$\Omega_\mathrm{L}$ is of order $t_{\rm LK}^{-1}$ at $e\sim 0$ (Equation \ref{eq:Omega L}), but increases with $e$. Depending on the
ratio of $\Omega_{\rm dS}$ and $\Omega_\mathrm{L}$, we expect three possible spin behaviors:
(i) For $\Omega_\mathrm{L}\gg \Omega_{\rm dS}$ (``nonadiabatic''),
the spin axis $\hat{\bf S}_1$ cannot ``keep up'' with the rapidly changing
$\hat{\textbf{L}}$, and thus effectively precesses around $\hat{\textbf{J}}$, keeping
$\theta_\mathrm{SJ}\equiv\cos^{-1}(\hat{\textbf{S}}_1\cdot\hat{\textbf{J}})\simeq$
constant.
(ii) For $\Omega_\mathrm{dS}\gg \Omega_\mathrm{L}$ (``adiabatic''),
$\hat{\textbf{S}}_1$ closely ``follows'' $\hat{\textbf{L}}$, maintaining an approximately constant
$\theta_\mathrm{SL}$.
(iii) For $\Omega_\mathrm{dS}\sim \Omega_\mathrm{L}$ (``trans-adiabatic''), the spin
evolution can be chaotic due to overlapping resonances.  Since both
$\Omega_\mathrm{dS}$ and $\Omega_\mathrm{L}$ depend on $e$ during the
LK cycles, the precise transitions between these regimes can be fuzzy
\citep{Dong Science,Storch 2015,Anderson et al HJ,Storch 2017}.

For circular orbits ($e=0$), the precession of $\hat{\bf L}$ is governed by the equation
\be\label{eq:simple L}
\frac{d \hat{\bf L}}{dt}\bigg|_{\lk,e=0}
=-\Omega_\mathrm{L}\hat{\bf L}_\OUT\times\hat{\bf L}
=-\Omega_\mathrm{L}'\hat{\bf J}\times\hat{\bf L},
\ee
where $\hat{\bf J}$ is the unit vector along ${\bf J}={\bf L}+{\bf L}_\OUT$, and
\be\label{eq:Omega L}
\Omega_\mathrm{L}=\frac{3}{4t_\lk}\big(\hat{\textbf{L}}\cdot\hat{\textbf{L}}_\OUT\big),
\quad
\Omega_\mathrm{L}'=\Omega_\mathrm{L}{J\over L_\OUT}.
\ee
We can define an ``adiabaticity parameter''
\be\label{eq:adiabaticity parameter}
\begin{split}
\mathcal{A}&\equiv \bigg(\frac{\Omega_\mathrm{dS}}{\Omega_\mathrm{L}}\bigg)_{e,I=0}
\simeq 0.37\bigg[\frac{(m_2+\mu/3)}{35M_\odot}\bigg]\bigg(\frac{m_{12}}{60M_\odot}\bigg)\\
&\times\bigg(\frac{m_3}{30M_\odot}\bigg)^{-1}
\bigg(\frac{a_{\OUT,\eff}}{3\au}\bigg)^3\bigg(\frac{a}{0.1\au}\bigg)^{-4}.
\end{split}
\ee
As the binary orbit decays, the system may transition from ``non-adiabatic''
(${\cal A}\ll 1$) at large $a$'s to ``adiabatic'' (${\cal A}\gg 1$) at small $a$'s, where
the final spin-orbit misalignment angle $\theta_{\rm SL}^{\rm f}$ is ``frozen''.
Note that ${\cal A}$ is directly related to $\varepsilon_\gr$ by
\be
{{\cal A}\over\varepsilon_\gr}={2\over 3}{m_2+\mu/3\over m_{12}}.
\label{eq:A-eps}\ee
Thus, when the initial value of $\varepsilon_\gr$ (at $a=a_0$)
satisfies $\varepsilon_{\gr,0}\lesssim 9/4$ (a necessary condition
for LK eccentricity excitation; see Equation~\ref{eq:grlim}), we also have
${\cal A}_0\lesssim (3m_2+\mu)/(2m_{12})\sim1$. This implies that any system that
experiences enhanced orbital decay due to LK oscillations
must go through the ``trans-adiabatic'' regime and therefore possibly chaotic spin evolution.

The bottom panel of Figure~\ref{fig:1} shows a representative example
of the evolution of the misalignment angle as the BH binary undergoes
LK-enhanced orbital decay. We see that the BH spin axis can exhibit
complex evolution even though the orbital evolution is ``regular''. In
particular, $\theta_\SL$ evolves in a chaotic way, with the final
value $\theta_\SL^{\rm f}$ depending sensitively on the initial
conditions (the precise initial $\theta_\SL$ and $I_0$). Also note
that retrograde spin ($\theta_\SL^{\rm f}>90^\circ$) can be produced
even though the binary always remains prograde with respect to the
outer companion ($I<90^\circ$). These behaviors are qualitatively similar to
the chaotic evolution of stellar spin driven by Newtonian spin-orbit coupling
with a giant planet undergoing high-eccentricity migration
\citep{Dong Science,Storch 2015,Anderson et al HJ,Storch 2017}.

We carry out a series of numerical integrations, evolving the orbit of
the merging BH binary with a tertiary companion, along with spin-orbit
coupling, to determine $\theta_\SL^\f$ for
various triple parameters. In our ``population synthesis'' study, we
consider initial conditions such that $\hat{\textbf{S}}_1$ is parallel to
$\hat{\textbf{L}}$, the binary inclinations are isotropically distributed
(uniform distribution in $\cos I_0$), and the orientations of
$\textbf{e}$ and $\textbf{e}_\OUT$ (for systems with $e_\OUT\neq 0$) are random.
All initial systems satisfy the criterion of dynamical stability for triples \citep{Mardling}.

\begin{figure}
\begin{centering}
\begin{tabular}{cc}
\includegraphics[width=7cm]{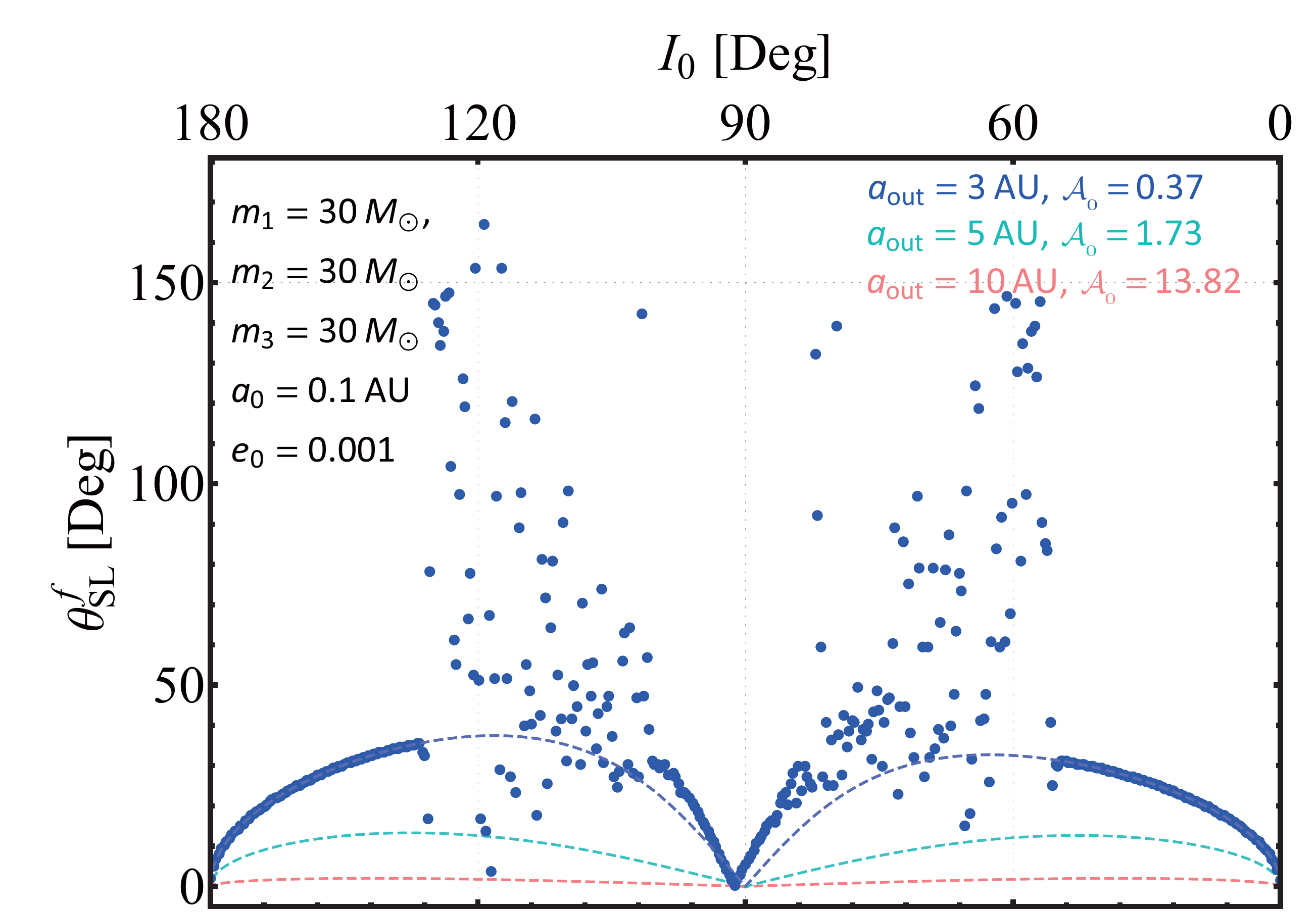}\\
\includegraphics[width=7cm]{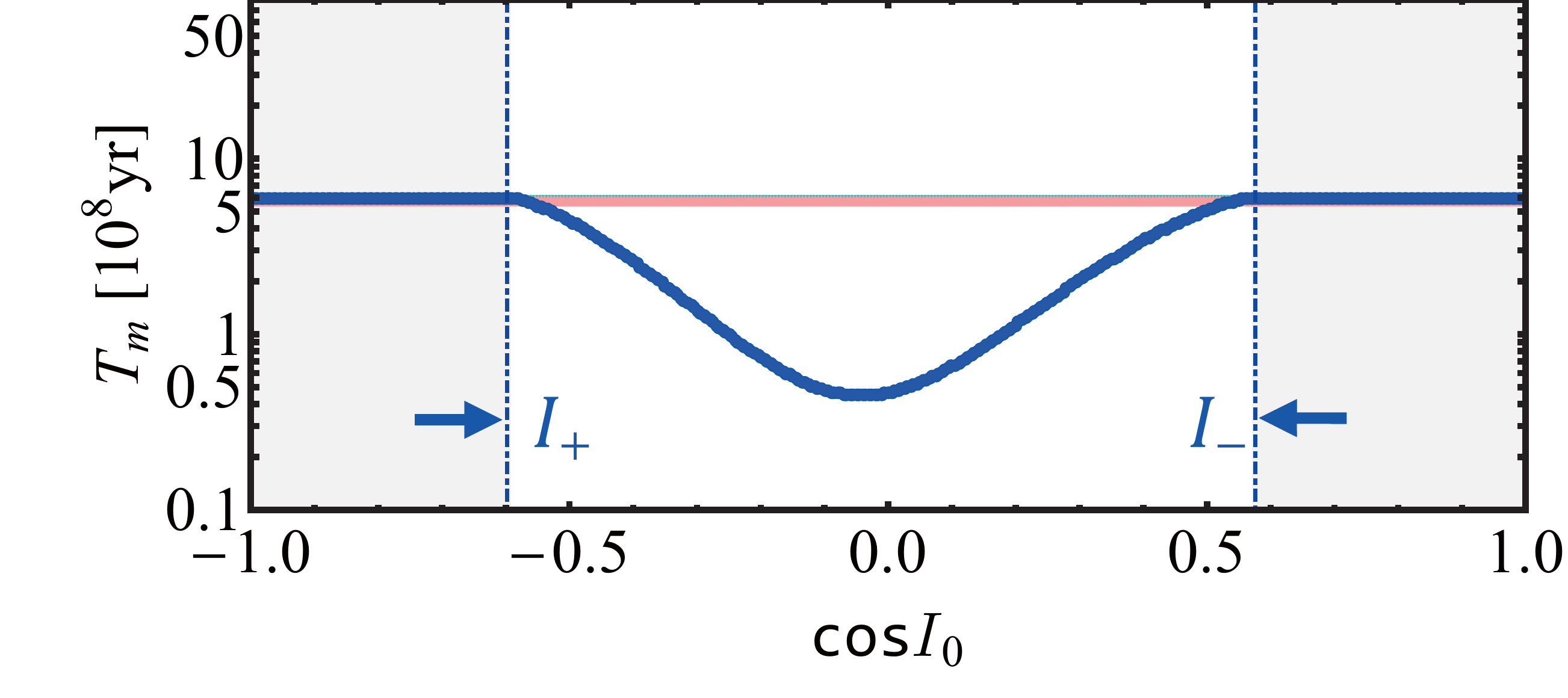}\\
\includegraphics[width=7cm]{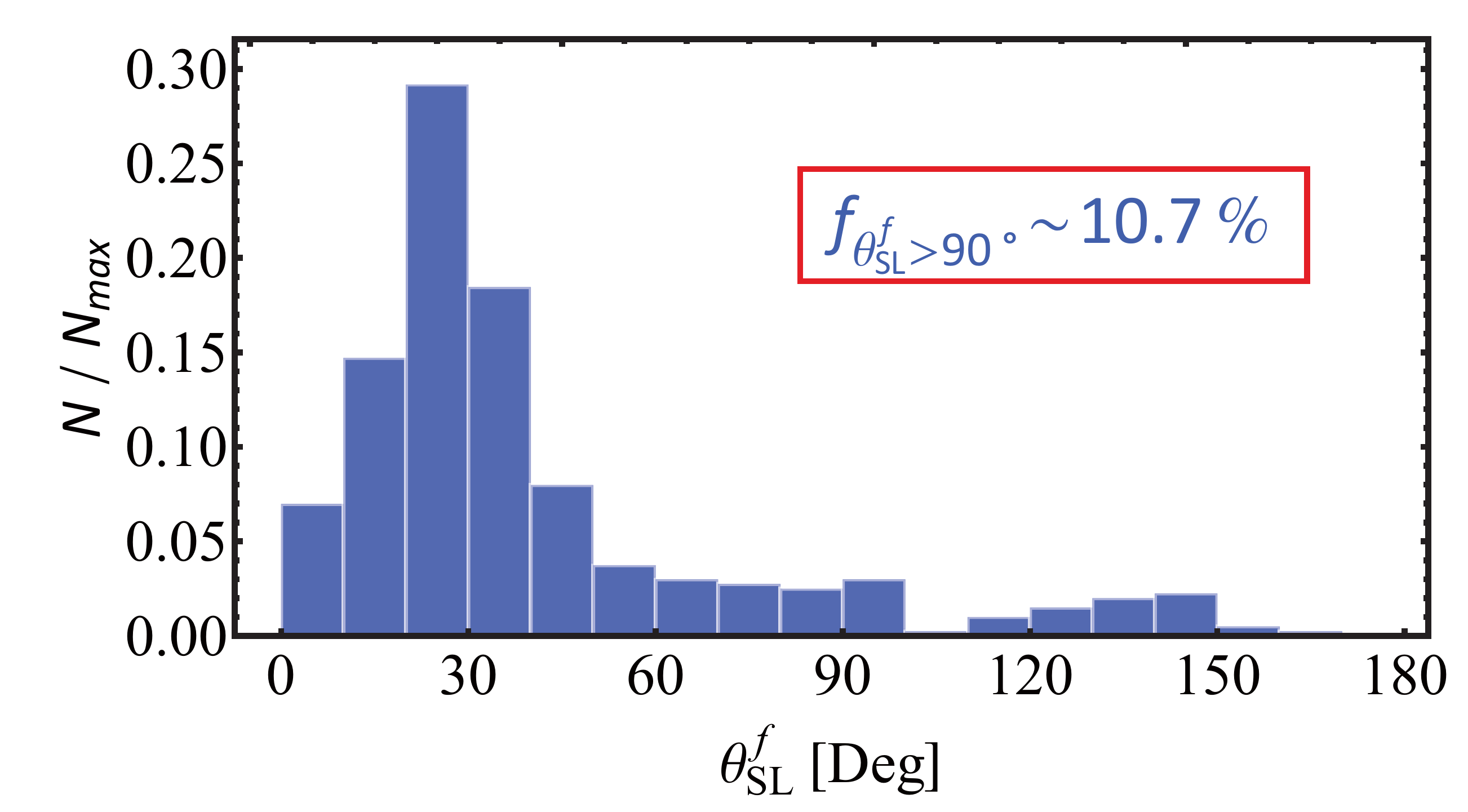}
\end{tabular}
\caption{Final spin-orbit misalignment angle (top panel) and merger
  time (middle panel) as a function of the initial inclination for
  equal mass triple systems with different outer semimajor axes (as
  labeled). The inner binary has fixed initial $a_0=0.1$~AU and $e_0=0.001$, and $e_\OUT=0$ for the outer binary.
In the top panel, the dots are the result of numerical integration for
the $a_\OUT=3$~AU system (a total of 400 runs on a uniform $\cos I_0$ grid),
 and the dashed curves are the analytical
results for circular orbits, as given by Equations~(\ref{eq:final misalignment}) and
(\ref{eq:tantheta}). The initial value of the adiabaticity
parameter $\mathcal{A}_0$ (Equation~\ref{eq:adiabaticity parameter} with $a=a_0$) is also given.
The vertical lines ($I_\pm$) shown in the middle panel
correspond to the LK window of eccentricity excitation
(Equation~\ref{eq:Kozai window I}).  The bottom panel shows the
distribution of the final spin-orbit misalignment angle for the system with
$a_\OUT=3\au$.  }
\label{fig:4}
\end{centering}
\end{figure}

Figure \ref{fig:4} shows our results for systems with equal masses,
$e_\OUT=0$ (so that the octupole effect vanishes), and several values of
$a_\OUT$. We see that when the eccentricity of the inner binary is excited
($I_0$ lies in the LK window), a wide range of
$\theta^\f_\SL$ is generated, including appreciable fraction of retrograde ($\theta_\SL^\f>90^\circ$)
systems (see the $a_\OUT=3$~AU case, for which $e_\li=0.84$). The ``memory'' of
chaotic spin evolution is evident, as slightly different initial
inclinations lead to vastly different $\theta_\SL^\f$. The regular behavior of
$\theta_\SL^\f$ around $I_0=90^\circ$ (again for the $a_\OUT=3$~AU case) is intriguing,
but may be understood using the theory developed in Ref.~\citep{Storch 2017}.
For systems with no eccentricity excitation, $\theta^\f_\SL$ varies
regularly as a function of $I_0$ -- this can be calculated analytically (see
below).

Figure \ref{fig:5} shows our results for systems with $m_1\neq m_2$
and $e_\OUT\neq 0$, for which octupole terms may affect the orbital evolution.
We see that eccentricity excitation and the corresponding reduction in $T_\mathrm{m}$ occur
outside the analytic (quadupole) LK window (see the $e_\OUT=0.8$ case, for which
$e_\li=0.66$). As in the equal-mass case (Figure~\ref{fig:4}), a wide
range of $\theta_\SL^\f$ values are produced whenever eccentricity excitation occurs.
A larger fraction ($23\%$) of systems attain retrograde spin
($\theta_\SL^\f > 90^\circ$). Again, for systems with negligible eccentricity
excitation, $\theta_\SL^\f$ behaves regularly as a function of $I_0$
and agrees with the analytic result (the ``fuzziness'' of the numerical result
in this regime is likely due to the very small eccentricity of the inner binary; see \citet{Liu-PRD}).

\begin{figure}
\begin{centering}
\begin{tabular}{cc}
\includegraphics[width=7cm]{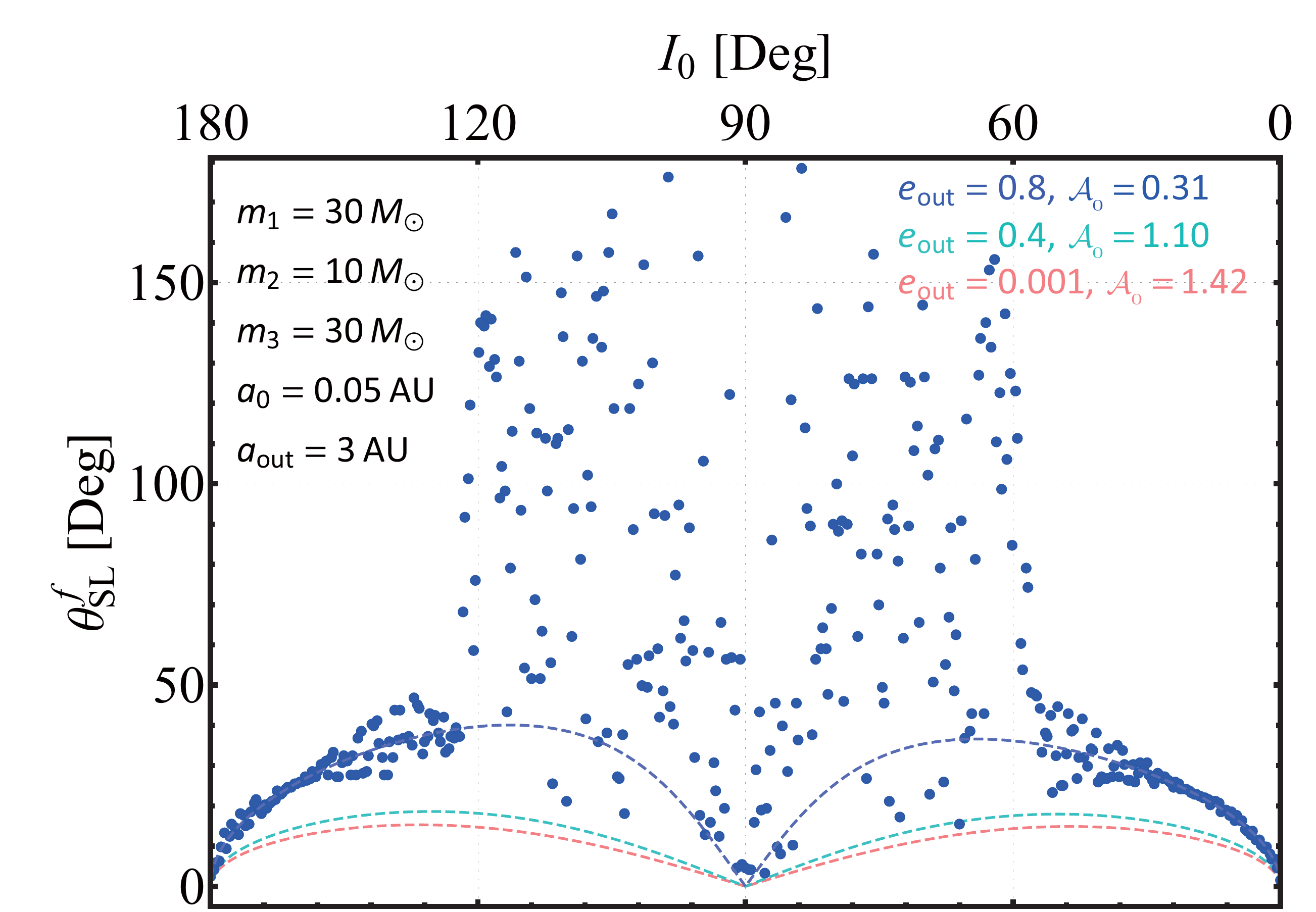}\\
\includegraphics[width=7cm]{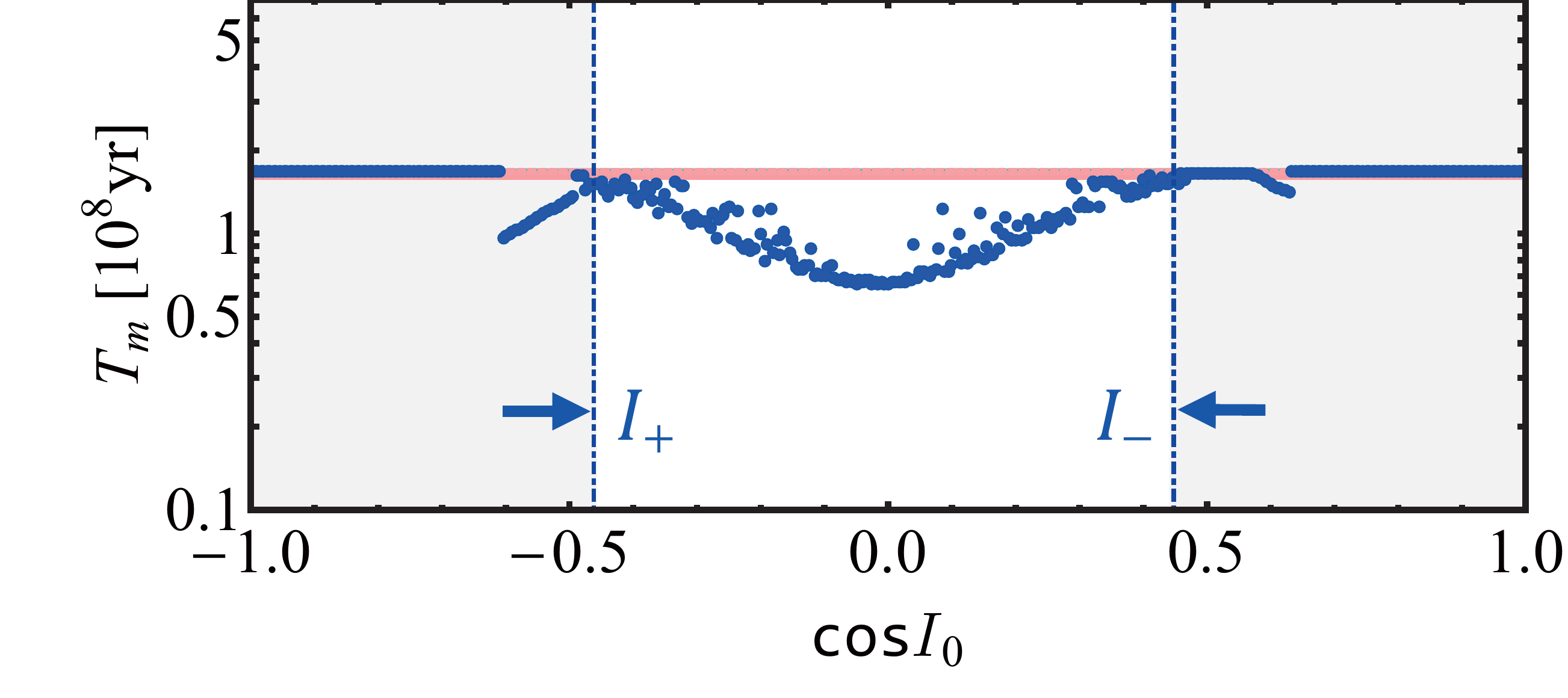}\\
\includegraphics[width=7cm]{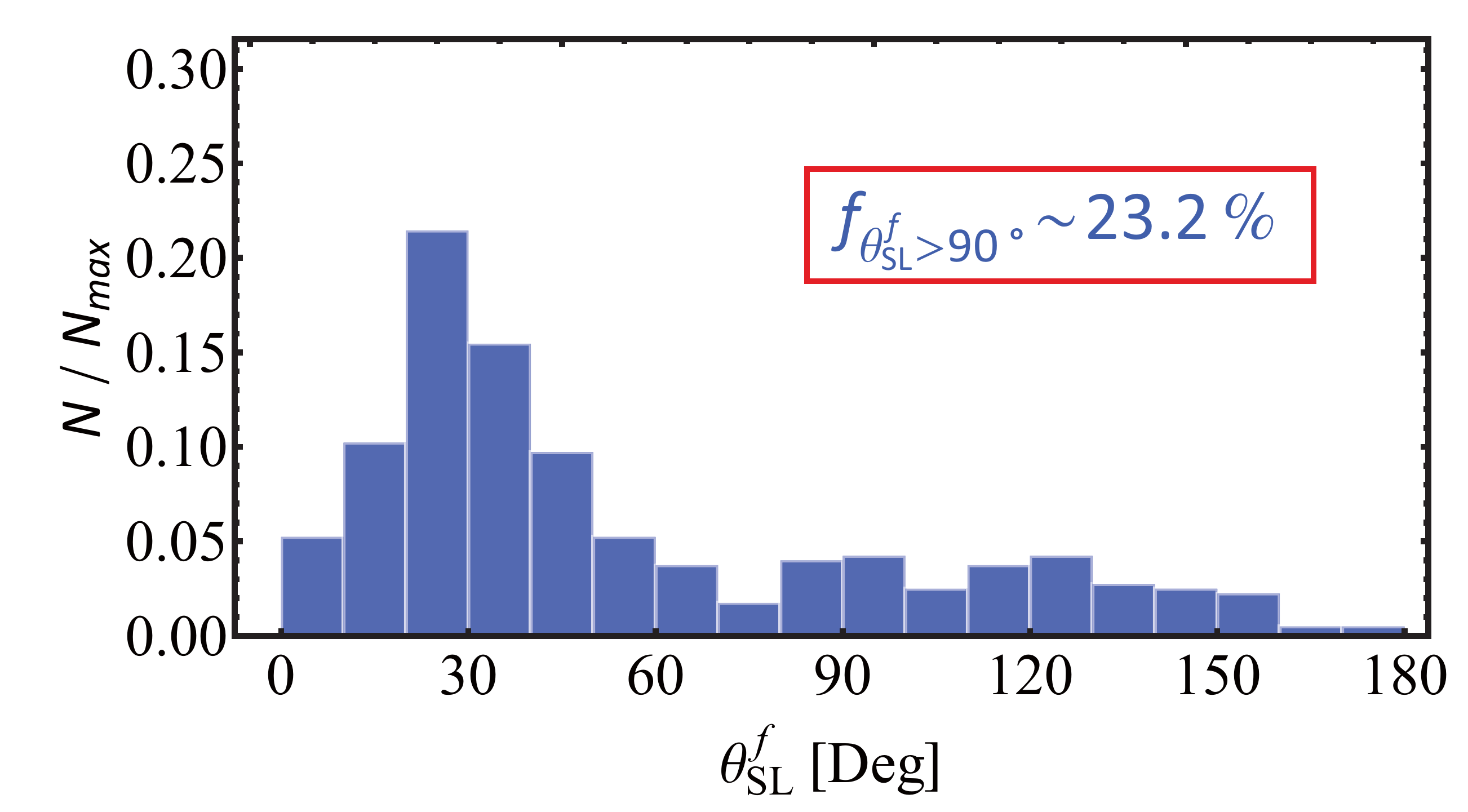}
\end{tabular}
\caption{Same as Figure\ref{fig:4}, but for unequal masses and non-circular outer orbits (as labeled).
The dots in the top panel are numerical results for the $e_\OUT=0.8$ system
(a total of 400 runs on a uniform $\cos I_0$ grid),
and the bottom panel shows the distribution of $\theta^\f_\SL$ for such a system.
}
\label{fig:5}
\end{centering}
\end{figure}

\subsection{Analytical Calculation of $\theta^\f_\SL$ for Circular Binaries}

If the inner binary experiences no eccentricity excitation and
remains circular throughout the orbital decay, the final spin-orbit misalignment
can be calculated analytically using the principle of adiabatic invariance.

Equation (\ref{eq:simple L}) shows that
$\hat{\bf L}$ rotates around the $\hat{\bf J}$ axis at the rate
$(-\Omega_\mathrm{L}')$. In this rotating frame, the spin evolution equation
(\ref{eq:spin}) transforms to
\be\label{eq:rotSpin}
\bigg(\frac{d\hat{\textbf{S}}_1}{dt}\bigg)_\mathrm{rot}=
\bm{\Omega}_\eff \times\hat{\textbf{S}}_1,
~~~\bm{\Omega}_\eff\equiv\Omega_\mathrm{dS}\hat{\textbf{L}}+\Omega'_\mathrm{L}\hat{\textbf{J}}.
\ee
Note that in the absence of GW dissipation, $\hat{\textbf{L}}$ and $\hat{\textbf{L}}_\OUT$
are constants (in the rotating frame), and thus $\hat{\bf S}_1$ precesses with a constant
$\bm{\Omega}_\eff$.
The relative inclination between $\bm{\Omega}_\eff$ and $\hat{\bf L}$ is given by
\be\label{eq:theta eff}
\tan\theta_{\eff,\mathrm{L}}=
\frac{\Omega_\mathrm{L}\sin I}{\Omega_\mathrm{dS}+\big(\eta+\cos I\big)\Omega_\mathrm{L}}.
\ee
Now if we include GW dissipation, $\hat{\textbf{L}}\cdot\hat{\textbf{L}}_\OUT=\cos I$ is exactly conserved,
and $\bm{\Omega}_\eff$ becomes a slowly changing vector.
When the rate of change of $\bm{\Omega}_\eff$ is much smaller than $|\bm{\Omega}_\eff|$,
the angle between $\bm{\Omega}_\eff$ and $\hat{\textbf{S}}_1$ is adiabatic invariant, i.e.
\be\label{eq:eff s}
\theta_{\eff,\mathrm{S}_1}\simeq \mathrm{constant}\qquad ({\rm adiabatic~invariant}).
\ee
This adiabatic invariance requires
$|d\bm{\Omega}_\eff/dt|/|\bm{\Omega}_\eff|\sim T_{m,0}^{-1}\ll |\bm{\Omega}_\eff|$, or
$|\bm{\Omega}_\eff|T_\mathrm{m,0}\gg1$, which is easily satisfied.

Suppose $\hat{\textbf{S}}_1$ and $\hat{\textbf{L}}$ are aligned initially,
we have $\theta_{\eff,\mathrm{S}_1}^0=\theta_{\eff,\mathrm{L}}^0$
(the superscript $0$ denotes initial value). Equation
(\ref{eq:eff s}) then implies
$\theta_{\eff,\mathrm{S}_1}\simeq \theta_{\eff,\mathrm{L}}^0$ at all times.
After the binary has decayed, $\eta\rightarrow 0$, $|\Omega_\mathrm{dS}|\gg|\Omega_\mathrm{L}|$,
and therefore $\bm{\Omega}_\eff\simeq\Omega_\mathrm{dS}\hat{\textbf{L}}$, which implies
$\theta_\SL^\f\simeq\theta_{\eff,\mathrm{S}_1}$.
Thus we find
\be\label{eq:final misalignment}
\theta_\SL^\f\simeq \theta_{\eff,\mathrm{L}}^0.
\ee
That is, the final spin-orbit misalignment angle is equal to the initial inclination angle between
$\hat{\bf L}$ and $\bm{\Omega}_\eff$, obtained by evaluating Equation~(\ref{eq:theta eff}) at $a=a_0$:
\be
\tan\theta_{\eff,\mathrm{L}}^0=
\frac{\sin I_0}{({\cal A}_0/\cos I_0) + \eta_0+\cos I_0}.
\label{eq:tantheta}\ee
This analytic expression agrees with the numerical results shown in Figures~\ref{fig:4}-\ref{fig:5}
in the appropriate regime.
Note that for systems that experience no eccentricity excitation for all $I_0$'s, ${\cal A}_0
\gtrsim (3m_2+\mu)/(2m_{12})\sim 1$ (see Equations~\ref{eq:grlim} and \ref{eq:A-eps}), and thus
$\theta_\SL^\f\simeq \theta_{\eff,\mathrm{L}}^0\lesssim 20^\circ$, i.e., only modest spin-orbit
misalignment can be generated. For systems with ${\cal A}_0\gg 1$ (e.g., very distant companion),
we have $\theta_\SL^\f\ll 1$.

The above analytical result can be easily generalized to the
situation of non-zero initial spin-orbit misalignment. It shows that
$\theta_\SL^\f\simeq \theta_\SL^0$ for ${\cal A}_0\gg 1$.

\section{Summary and Discussion}

We have studied the effect of external
companion on the orbital and spin evolution of merging BH binaries due
to gravitational radiation.
A sufficiently close by and inclined companion can excite Lidov-Kozai eccentricity oscillation
in the binary, shortening its merger time compared to circular orbits
[see Fig.\ref{fig:3}].  We find that during
the LK-enhanced orbital decay, the spin axis of the BH generally
experiences complex, chaotic evolution, with the final spin-orbit
misalignment angle $\theta_\SL^\f$ depending sensitively on the initial
conditions. A wide range of $\theta_\SL^\f$ (including
$\theta_\SL^\f>90^\circ$) can be produced from an initially aligned
($\theta_\SL^0=0$) configuration (see Figs.\ref{fig:4}-\ref{fig:5}). For
systems that do not experience eccentricity excitation (because of
relatively low orbital inclinations of the companion or/and suppression by GR-induced
precession), modest ($\lesssim 20^\circ$) spin-orbit misalignment can be produced -- we
have derived an analytic expression for $\theta_\SL^\f$ for such
systems (Eqs.\ref{eq:final misalignment}-\ref{eq:tantheta}).  Note
that while our numerical results refer to stellar-mass
companions, our analysis is not restricted to any specific binary formation
scenarios, and can be easily adapted to other types of systems (e.g. when the tertiary is a
supermassive BH) by applying appropriate scaling relations.
The key dimensionless parameter that determines the spin-orbit evolution is
${\cal A}_0$  (see Eq.\ref{eq:adiabaticity parameter}).

The BH binaries detected by aLIGO so far \citep{Abbott 2016b,Abbott 2017}
have relatively small $\chi_{\rm eff}$ ($0.06^{+0.14}_{-0.14}$ for GW150914,
$0.21^{+0.2}_{-0.1}$ for GW151226, and $-0.12^{+0.21}_{-0.30}$ for GW170104).
These small values could be due to the slow rotation of the BHs \citep{Zaldarriaga} or
spin-orbit misalignments. The latter possibility would imply a dynamical formation channel
of the BH binaries (such as exchange interaction in globular clusters \citep{Rodriguez 2015,Chatterjee 2017})
or, as our calculations indicate, dynamical influences of external companions.

BL thanks Natalia Storch, Feng Yuan and
Xing-Hao Liao for discussions. This work is supported in
part by grants from the National Postdoctoral Program and NSFC (No.BX201600179, No.2016M601673 and No.11703068).
DL is supported by NASA grants NNX14AG94G and NNX14AP31G,
and a Simons Fellowship in theoretical physics.
This work made use of the Resource in the Core
Facility for Advanced Research Computing at SHAO.



\begin{thebibliography}{}

\bibitem[Abbott et al.(2016a)]{Abbott 2016a} Abbott, B. P., et al. (LIGO Scientific and Virgo Collaboration) \ 2016a, PhRvL, 116, 061102

\bibitem[Abbott et al.(2016b)]{Abbott 2016b} Abbott, B. P., et al. (LIGO Scientific and Virgo Collaboration) \ 2016b, PhRvX, 6, 041015

\bibitem[Abbott et al.(2017)]{Abbott 2017} Abbott, B. P., et al. (LIGO Scientific and Virgo Collaboration) \ 2017, PhRvL, 118, 221101

\bibitem[Anderson et al.(2016)]{Anderson et al HJ} Anderson, K. R., Storch, N. I., \& Lai, D. \ 2016, MNRAS, 456, 3671

\bibitem[Anderson et al.(2017a)]{Anderson et al 2017} Anderson, K.~R., Lai, D., \& Storch, N.~I. \ 2017, MNRAS, 467, 3066

\bibitem[Anderson \& Lai(2017b)]{Anderson 2017b} Anderson, K.~R., \& Lai, D. \ 2017, arXiv:1706.00084

\bibitem[Antonini \& Perets(2012)]{Antonini 2012} Antonini, F., \& Perets, H. B. \ 2012, ApJ, 757, 27

\bibitem[Antonini et al.(2014)]{Antonini 2014} Antonini, F., Murray, N., \& Mikkola, S. \ 2014, ApJ, 781, 45

\bibitem[Antonini \& Rasio(2016)]{Antonini and Rasio 2016} Antonini, F., \& Rasio, F.~A. \ 2016, ApJ, 831, 187

\bibitem[Antonini et al.(2017)]{Antonini 2017} Antonini, F., Toonen, S., \& Hamers, A. S. \ 2017, ApJ, 841, 77

\bibitem[Barker \& O'Connell(1975)]{Barker} Barker, B.~M., \& O'Connell, R.~F. \ 1975, PhRvD, 12, 329

\bibitem[Belczynski et al.(2010)]{Belczynski 2010} Belczynski, K., Benacquista, M., \& Bulik, T. \ 2010, ApJ, 725, 816

\bibitem[Belczynski et al.(2016)]{Belczynski 2016} Belczynski, K., Holz, D.~E., Bulik, T., \& O'Shaughnessy, R. \ 2016, Nature, 534, 512

\bibitem[Blaes et al.(2002)]{Blaes 2002} Blaes, O., Lee, M.~H., \& Socrates, A. \ 2002, ApJ, 578, 775

\bibitem[Chatterjee et al.(2017)]{Chatterjee 2017} Chatterjee, S., Rodriguez, C.~L., Kalogera, V., \& Rasio, F.~A. \ 2017, ApJL, 836, L26

\bibitem[Ford et al.(2000)]{Ford} Ford, E. B., Kozinsky, B., \& Rasio, F. A. \ 2000, ApJ, 535, 385

\bibitem[Kozai(1962)]{Kozai} Kozai, Y. \ 1962, AJ, 67, 591

\bibitem[Lidov(1962)]{Lidov} Lidov, M. L. \ 1962, Planetary and Space Science, 9, 719

\bibitem[Lipunov et al.(1997)]{Lipunov 1997} Lipunov, V. M., Postnov, K. A., \& Prokhorov, M. E \ 1997, AstL, 23, 492

\bibitem[Lipunov et al.(2017)]{Lipunov 2017} Lipunov, V. M., et al. \ 2017, MNRAS, 465, 3656

\bibitem[Liu et al.(2015a)]{Liu et al 2015} Liu, B., Mu{\~n}oz, D.~J., \& Lai, D. \ 2015a, MNRAS, 447, 747

\bibitem[Liu et al.(2015b)]{Liu-PRD} Liu, B., Lai, D., \& Yuan, Y.-F. \ 2015b, PhRvD, 92, 124048

\bibitem[Mandel \& de Mink(2016)]{Mandel and de Mink 2016} Mandel, I., \& de Mink, S.~E. \ 2016, MNRAS, 458, 2634

\bibitem[Mardling \& Aarseth(2001)]{Mardling} Mardling, R. A., \& Aarseth, S. J. \ 2001, MNRAS, 321, 398

\bibitem[Miller \& Hamilton(2002)]{Miller 2002} Miller, M. C., \& Hamilton, D. P. \ 2002, ApJ, 576, 894

\bibitem[Naoz(2016)]{Naoz 2016} Naoz, S. \ 2016, ARA\&A, 54, 441

\bibitem[Petrovich \& Antonini(2017)]{Petrovich 2017} Petrovich, C., \& Antonini, F. \ 2017, arXiv:1705.05848

\bibitem[Rodriguez et al.(2015)]{Rodriguez 2015} Rodriguez, C.~L., et al. \ 2015, PhRvL, 115, 051101

\bibitem[Shapiro \& Teukolsky(1983)]{Shapiro 1983} Shapiro, S. L., \& Teukolsky, S. A. \ 1983,
Black Holes, White Dwarfs, and Neutron Stars: The Physics of Compact Objects, (Wiley, New York, 1983)

\bibitem[Silsbee \& Tremaine(2017)]{Silsbee and Tremaine 2017} Silsbee, K., \& Tremaine, S. \ 2017, ApJ, 836, 39

\bibitem[Storch et al.(2014)]{Dong Science} Storch, N. I., Anderson, K. R., \& Lai, D. \ 2014, Science 345, 1317

\bibitem[Storch \& Lai(2015)]{Storch 2015} Storch, N. I., \& Lai, D. \ 2015, MNRAS, 448, 1821

\bibitem[Storch et al.(2017)]{Storch 2017} Storch, N. I., Lai, D., \& Anderson, K. R. \ 2017, MNRAS, 465, 3927

\bibitem[Thompson(2011)]{Thompson 2011} Thompson, T.~A. \ 2011, ApJ, 741, 82

\bibitem[Zaldarriaga et al.(2017)]{Zaldarriaga} Zaldarriaga, M., Kushnir, D., \& Kollmeier, J. A. \ 2017, arXiv:1702.00885

\end{thebibliography}
\end{document}